# A First Implementation of a Design Thinking Workshop During a Mobile App Development Project Course

Yen Dieu Pham
University of Hamburg
Hamburg, Germany
pham@informatik.uni-hamburg.de

Davide Fucci
University of Hamburg
Hamburg, Germany
fucci@informatik.uni-hamburg.de

Walid Maalej
University of Hamburg
Hamburg, Germany
maalej@informatik.uni-hamburg.de

## ABSTRACT

Due to their characteristics, millennials prefer learning-by-doing and social learning, such as project-based learning. However, software development projects require not only technical skills but also creativity—Design Thinking can serve such purpose. We conducted a workshop— following the Design Thinking approach of the *d.school*—to help students generating ideas for a mobile app development project course. On top of the details for implementing the workshop, we report our observations, lessons learned, and provide suggestions for further implementation.

## KEYWORDS

software engineering education, design thinking, project-based learning, team-based learning.

## 1 INTRODUCTION

Millennials—the tech-savvy generation of university students—prefer learning-by-doing, group works, and the social aspects of learning [1]. For them, the roles of consumer and producer of creative work are blending together [2].

Those features need to be considered when educating students in software engineering—a field which requires practice and where theory is not sufficient to solve real, professional-grade problems. That is why the software engineering curricula often include project-based courses, in which students have the chance to apply traversal skills (e.g., software development, project management, UX design) while addressing a complex, possibly real-world problem.

In this context, rather than focusing on specific, technical knowledge (e.g., software development), teachers should support the students' learning process and help them to deal with complex problems while exploring diverse solutions. This approach is not static and requires creative problem solving [3].

In the last few years, Design Thinking has emerged as a problem-solving approach for "wicked problems", settings which are characterized by incomplete, contradictory, ambiguous, and changing requirements [4]. Design Thinking supports the generation of ideas and solutions (e.g., products, services) which are "viable and novel for a particular group of users" [5]. It is being used as a teaching approach—particularly, in co-location with project-based learning—not only by design schools but also in other disciplines [6], including engineering [7]. In this paper, we report on our experience of running a design thinking workshop during *M-Lab*—a project-, team-based course focusing on app development which involves industrial customers at the University of Hamburg. The workshop was implemented as an intervention for those teams struggling with formulating a viable solution for the problems presented by their customers. Such scenario was ideal for implementing Design Thinking and engage the students to generate creative solutions. In particular, we used the Design Thinking approach proposed by the d.school (Section 2) and followed several of their suggested methods during the workshop (Section 3). In this paper, we report our lesson learned and suggest further improvement to the proposed approach (Section 4).

## 2 DESIGN THINKING MATERIAL AND METHODS FROM d.school

In this section, we define the Design Thinking approach we utilized during the workshop based on the methods developed by the Hasso Plattner Institute of Design (also known as d.school) founded at *Stanford University* in 2005. We decided to use this specific approach as it has already been tested in an engineering environment rather than a pure design approach [8].

Although there is no generally accepted definition [8] Design Thinking can be understood as a framework with a *"human-centered approach to problem solving"* [9]. Depending on the context, Design Thinking can also be interpreted as an innovation method, a working procedure, an attitude towards life, a mindset, or a tool [8].

d.school defines *Design Thinking* approach as a "constant-work-in-progress" framework of working modes (Section 2.1) and mindsets (Section 2.2) [10].

### 2.1 d.school WORKING MODES

The working mode are process phases, which consists of five iterative steps (see Figure 1), described as follows [10]:

**Figure 1: Working Modes from d.school [11].**

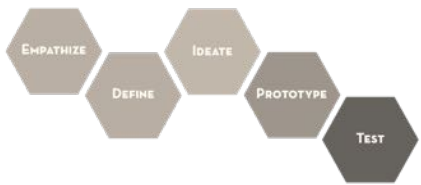

**Empathize.** In this mode, the designers should try to understand their users. The theory is that the problems of the users are not often related to the designers. Designers need to empathize with the users to design appropriate solutions that fit their needs.

**Define.** In this mode, the designers need a deep understanding of their users to identify their problems. Based on these insights, designers can "scope a specific and meaningful challenge".

**Ideate.** The challenge defined in the previous phase represents the starting point for the designers to look for a solution. This process supports the designers in generating new ideas, they will have an explicit problem to solve and know where to start. Moreover, while searching for a solution, the designers will come up with new ideas.

**Prototype.** A prototype is defined as "anything that takes a physical form" [10]. It is useful to test functionalities, deepen the users understanding, inspire teammates, and explore more solutions.

**Test.** The testing mode gives the designer the opportunity to get feedback from the users. The latter will improve the idea and lead to new insights.

## 2.2 d.school DESIGN THINKING MINDSETS

In the guide "The Bootcamp Bootleg"[1], the d.school states seven mindsets (reported in Table 1)—a toolkit to support the design thinking practice. The mindsets describe the attitude designers need to practice Design Thinking.

**Table 1: Mindsets from the d.school "The Bootcamp Bootleg".**

| MINDSET | DESCRIPTION |
| --- | --- |
| Show, don't tell | Communicate your vision in an impactful and meaningful way by creating experience, using illustrative visuals and telling good stories. |
| Focus on human values | Empathy for the people you are designing for and feedback from these users is fundamental to good design. |
| Be mindful of the process | Know where you are in the design process, the methods to use in that stage, and what are your goals. |
| Bias towards action | The name Design *Thinking* is a misnomer; it is more about doing than thinking. Be biased toward doing and making over thinking and meeting. |
| Radical collaboration | Bring together innovators with diverse backgrounds and viewpoints. Enable breakthrough insights and solutions to emerge from diversity. |
| Embrace experimentation | Prototyping is not simply a way to validate your idea, but an integral part of the innovation process. Prototypes are built to think and learn |
| Craft clarity | Produce a coherent vision out of complex problems. The problem needs to be framed to inspire others and fuel ideation. |

## 3 IMPLEMENTING THE d.school DESIGN THINKING DURING M-LAB

### 3.1 WORKSHOP MOTIVATION

The d.school Design Thinking approach was implemented as an intervention workshop within *M-Lab*[2]—a semester-long project-based course at the department of Informatics, University of Hamburg (Germany). During the project, each of the five teams, consisting of five to eleven students, develop a mobile app for a real customer (e.g., the local university hospital). According to the syllabus of the project, the students should have generated "innovative" ideas (to be later implemented) two months after the beginning of the course. Two teams were struggling with idea/solution generation; hence, the teaching staff decided to intervene with a Design Thinking workshop. A mix of 11 Bachelor and Master students—four from the software engineering curricula, five from Information Systems, one from Software System Development, and one from Human-Computer Interaction—attended the workshop. The teaching was interviewed before the workshop; they reported that the students had issues coming up with innovative ideas, and that rather than trying to create something new or remarkable, they were more concerned with meeting all the formal criteria to pass the course (e.g., writing a problem statement, developing a clickable prototype). According to the teaching staff, the team facing the most difficulties was the one that did not get any specific requirements from their customer, a telecommunication provider—i.e., the customer gave them complete freedom, as long as they would deliver an *innovative* mobile app. Instead of being inspired by the possibilities, the students felt overstrained and clueless. Based on these insights, we wanted to achieve two types of goals, internal (i.e., not communicated to the students) and external (i.e., presented to the students).

---

[1] https://dschool.stanford.edu/resources/the-bootcamp-bootleg

[2] https://mast.informatik.uni-hamburg.de/mlab

**Internal Goals**
- Inspire creativity to the students
- Improve their confidence and enable them to come up with ideas in a limited amount of time
- Help the group struggling the most, without exposing them as "weak" to their peers
- Provide each team with valuable experiences from their own progress

**External Goals**
- Reflect on the current state of their project
- Create new ideas as well as concrete suggestions for their implementation

### 3.1 WORKSHOP IMPLEMENTATION

We summarize the implementation of the workshop in Table 4, focusing on:
- Process or activities carried out by the facilitator and the students.
- Goals to be achieved by the facilitator and students.
- Materials used by the facilitator for the preparation and during the workshop.
- Formation (e.g., grouping) of the students during the different steps.
- Time frame
- Corresponding d.school Working mode

The workshop took place at the University of Hamburg. It was implemented in a room with space for 30 people (see Figure 2) The materials used were pinboards, a flip chart, Post-It notes, markers, A3 and A0 sheets A.

**Figure 2: World Café Round 3**

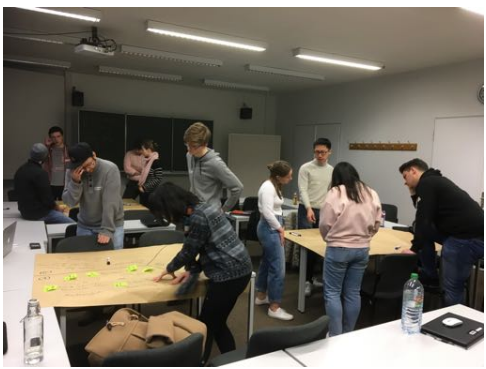

The first author of this paper acted as the facilitator and put the workshop into practice. The workshop lasted for two hours and consisted of five steps (see Figure 3):

**Figure 3: Adjusted Working Modes**

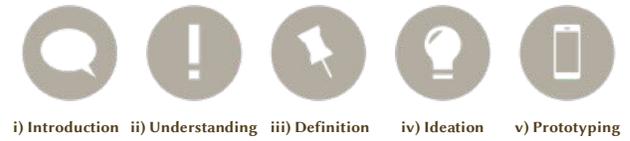

i) Introduction   ii) Understanding   iii) Definition   iv) Ideation   v) Prototyping

**i) Introduction:**
The facilitator and students introduce themselves and tell each other about their backgrounds.

**ii) Understanding:**
Every team reflects their previous progress with a template (see Figure 4) and should realize if they understood and covered the needs and problems of their users. This phase should be used as a starting point to build up empathy for the users.

**Figure 4: "Understanding" Template Customer/ Users**

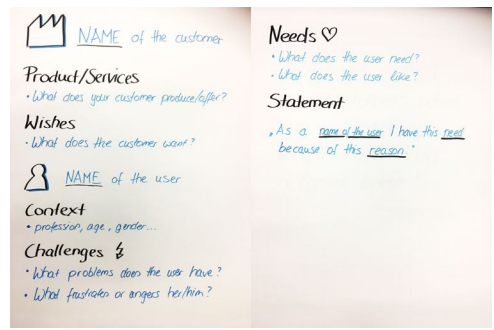

**Transition** from Understanding to Definition:
The facilitator chooses an app project in agreement with the students. In accordance with the teaching staff the students will focus on the telecommunications app.

**iii) Definition:**
The facilitator collects the experiences of the students as smartphone users and customers of a telecommunications provider. The facilitator closes this phase with a clear scope which means to focus on three.

**2. Transition** from Definition to Ideation
The facilitator prepares students to develop an open-minded mindset with an adjusted version of the improvisation game "Yes, but…/Yes, then let's…" (reported in Table 2)

Table 2: Instructions of Improvisation Game "Yes, but.../ Yes and then let's..."

| PHASE | ACTION |
|---|---|
| Preparations | - Each student chooses a partner whom she or he never really interact before.<br>- The students should get used to it to work with unknown partners. |
| Round 1:<br>"Yes, but..." | - Student A thinks about a destination, where she or he would like to travel<br>- Student B has to find arguments against the ideas of Student A and should start her/his sentences with: "Yes, but...<br>- The dialogue should last about 2-3 minutes:<br>- Example of a dialogue:<br>  - Student A: "Let's travel to Hawaii"<br>  - Student B: "Yes, but it is so far away"<br>  - Student A: "Maybe, but it is sunny and we can relax at the beach."<br>  - Student B: "Yes, but I will probably get a sunburn and have to stay in the shadow the rest of the journey"<br>  - Student A: ... |
| Round 2:<br>"Yes, and then let's..." | - Student A starts with the same destination<br>- Student B has to build on the ideas of Student A, and should start her/his sentence with: "Yes, and then let's..."<br>- The dialogue should last about 2-3 minutes:<br>- Example of a dialogue:<br>  - Student A: "Let's travel to Hawaii"<br>  - Student B: "Yes and then let's spend some time on the beach"<br>  - Student A: "Exactly and go swimming"<br>  - Student B: "Yes and then let's rent a boat to visit all the small islands of Hawaii" |

**iv) Ideation**

The students create ideas together with the "World Café" method reported in Table 3. During each round they fill in the template shown in Figure 5.

**Figure 5: World Café Template**

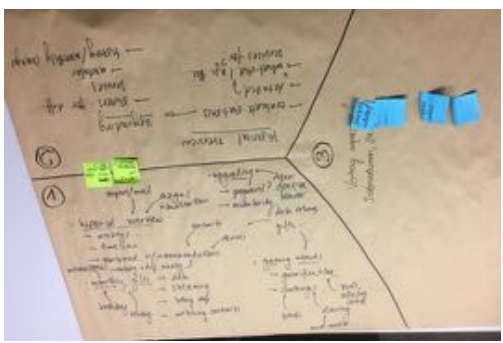

Table 3: World Café Procedure

| PHASE | ACTION |
|---|---|
| Preparations | - The students spread equally according to the number of tables |
| Round 1: | - Each table team creates ideas or solutions based on the topics of the "Definition" phase |
| Transition to Round 2 | - Each team chooses a "table master"<br>- The "table master" has to stay<br>- The other students spread individually to a new topic/ table |
| Round 2: | - The "table master" presents the results of the previous group to the new members<br>- The new team creates ideas or solutions based on the results of the previous group. |
| Transition to Round 3 | - The first "table master" has to leave the table and the team decides on a new one.<br>- The other students spread individually to a new table again. |
| Round 3 (see Figure 4): | - The new "table master" presents the previous ideas to the new team<br>- Last round of ideation based on the ideas of the previous teams. |

**v) Prototyping**

The students transform their theoretical ideas into paper prototypes (see Figure 6) These prototypes should serve as inspirations and suggestions how to implement the ideas.

**Figure 6: Paper Prototypes Round 3**

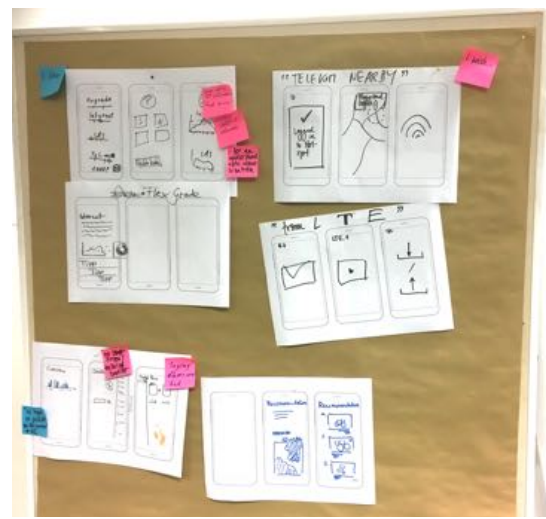

**Debriefing**

Summary of the results, workshops and learnings. The facilitator asks for feedback from the students.

**Table 4: Detailed Description of the implemented Design Thinking (DT) Workshop.**

| PROCESS | GOALS | MATERIALS | FORMATION | TIME | DT MODE |
|---|---|---|---|---|---|
| 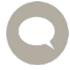 **i) Introduction** ||||||
| • Facilitator and students introduce themselves (name, background, app-team).<br>• Flipchart presentation of the goals and schedule (displayed during the entire workshop duration). | • Creating a trustful atmosphere<br>• Involve students, so that they are more likely to express their ideas by [12]:<br>  - Knowing each other's name and background.<br>  - Being transparent about the procedure. | • Tape, name tag for the participants<br>• Flipchart | • 1 Group, 11 students 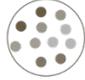 | • 5 Min. | |
| 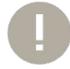 **ii) Understanding** ||||||
| • Facilitator presents a template with questions about their customer and users (see Figure 4). | • The students should recap their current progress and what they might have missed. | • Flipchart | • 5 Groups, divided by appprojects 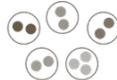 | • 10 min. (brainstorming)<br>• 10 min. (presentation) | • Empha-size |
| → *TRANSITION TO THE DEFINING PHASE:* ||||||
| • In agreement with the students, the facilitator decides to focus on the telecommunication team | • Focusing on the telecommunication team without putting them in a difficult spot.<br>• Creating a clear scope. | | • 1 Group, 11 students 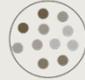 | • 5 Min.<br>*1* | • Empha-size → Define |
| 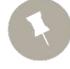 **iii) Defining** ||||||
| • The facilitator<br>  - asks the students about their experiences with their telecommunication provider.<br>  - writes down each experience on a Post-It. | • Collecting needs and challenges, which should be inspirations for solutions/new ideas. | • A0 sheets<br>• Pinboard<br>• Post-it | • 1 Group, 11 students 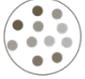 | • 15 min. (work) | • Define |
| • 2. Clustering.<br>• The facilitator arranges Post-It in clusters on the board while brainstorming with them. | • Getting an overview of the topics and preparing for the voting. | | | | |
| • 3. Voting:<br>• Each student votes for their most interesting topic/cluster | • Defining a clear goal. | | | | |
| → *TRANSITION TO THE IDEATION PHASE:* ||||||
| • Improvisation Game:<br>  - Yes, but… / Yes and then… (see Table 2) | • Making students realize how important it is to be open-minded and supportive towards to teammates to create new ideas [12]. | - | • 6 pairs 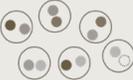 | • 5 min. | • Define → Ideate |
| 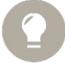 iv) Ideation | | | | | |
| • World Café [13] (a technique for generating ideas—see Table 3):<br>  - Facilitator puts the three most voted topics on different tables with a template (Figure 5).<br>  - Students dived equally (max. four students) among the tables.<br>  - Students rotate topics and teammates three times. | • Creating ideas based on other people ideas.<br>• Collaboration with different people.<br>• Students realize that too much discussion in the ideation phase prevents new ideas. | • A0 sheets<br>• Tables for 3 A0 sheets | • 3 rotating groups 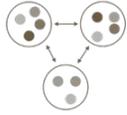 | • 30 min. (each round 10 min.) | • Ideate |
| 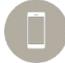 v) Prototyping | | | | | |
| • First round of prototyping:<br>  - Students stay in their last World Café.<br>  - The facilitator hands out a paper with smartphone outlines (see Figure 6).<br>  - The students have to sharpen their solutions by building a prototype. | • Creating suggestions and inspiration for concrete implementations.<br>• | • A3 Sheets with Smart-phone Outlines | • 3 groups 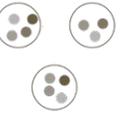 | • 10 min. | • Prototype |
| • First feedback with method "I wish/I like":<br>  - Teams present their prototypes<br>  - Students comment on prototype what they like/what they wish the prototype should offer or change.<br>  - Facilitator writes down the feedback. | • Getting constructive feedback from the other groups.<br>• | • Post-It | • 1 group, 11 students 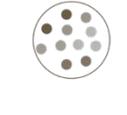 | • 15 min., each group 5 min. for presentation and feedback | • Test |
| • Second round of prototyping:<br>  - Adapting the feedback<br>  - Presentation. | • Realizing that early feedback is very helpful. | • A3 Sheets with Smart-phone Outlines | • 3 groups 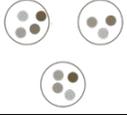 | • 10 min. | • Prototype |
| • Second round of feedback<br>  - Same as first feedback. | • Getting constructive feedback from the other groups.<br>• | • Post-It | • 1 group, 11 students 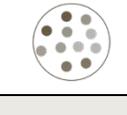 | • 15 min. | |
| → *DEBRIEFING* | | | | | |
| • Facilitator sums up the results of the workshop. | • Building up students' self-confidence by showing that they did a lot in a short amount of time.<br>• Convincing students that their mindset is important to be creative. | • All the results on Pinboard<br>• Flipchart | • 1 group, 11 students 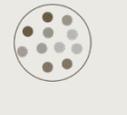 | • 3-5 min. | |
| • Feedback:<br>• The facilitator gives students a small ball:<br>  - The students toss the ball to one another.<br>  - The student with the ball should give a feedback. | • Collecting suggestions to improve future workshops. | | | • 10 min. | |

# 4 RESULTS

The workshop had the effect of changing the students' perspective towards a human-centred approach. At last, their final ideas were based on a need of a potential user.

From the educator perspective, the biggest challenge was deciding on methods in line with:
- The internal and external goals,
- The mindsets of the d.school Design Thinking,
- The available amount of time.

This section reports i) our most salient observations gathered during the workshop; ii) the associated lessons learned; iii) and suggestions for further iterations. Moreover, we report our appraisal after the completion of the M-Lab course.

## 4.1 DURING THE WORKSHOP.

We divide our observations, lessons learned, and challenges according to the specific steps.

**Understanding phase:**
*Observation.* The facilitator observed that the description of the template (see Figure 4) was not specific enough. In particular, some students described their users in general, other described a personification of the user. The students were confused by the term "name", which meant for some groups a real name like "Jane" and not the general name of their user group.
In fact, the facilitator wanted them to take the perspective of their typical user group rather than making up a persona. The students should look at a group to get a wide overview of needs and problems. A persona can be too narrow and misleading if the students did not interview their customers and final users.
*Lessons learned 1.* Being more careful with ambiguous description of the material.
*Suggestions 1.* The template should explicitly state the characterization of the users in the *Understanding template* (see Figure 3). For example, express the naming of the user group as an activity rather than a noun.

**Defining phase:**
*Observation.* At the beginning, we were worried that the students would not engage with the workshop and could not develop needs and problems because their academic background was too similar in contrast with the mindset of "*radical collaboration*". Although the facilitator asked about the students experiences and background, three did not provide any. Nevertheless, it was a productive phase, thanks to the input from the students with a more creative background (e.g., in HCI/UX) who provided half of the discussion topics.
*Lessons learned 2.* Not every student is used to express her or his ideas openly nor able to develop them in a group.
*Suggestions 2.* Before starting the open discussion, every student should get few minutes to think and develop their ideas individually.

**Ideation phase:**
*Observation.* During the "World Café", some groups took too long to start sketching their ideas. The facilitator had to refrain the students from discussing too much by reminding them about the time limit.
*Lessons learned 3.* It was challenging for the facilitator to keep an eye on every single group, as well as tracking time and to make sure that the groups would not get stuck discussing details. Therefore, some part of the management of the group should be done by the students themselves.
*Suggestions 3.* At the beginning of each round, the students should assign roles, like "writer" and "time keeper" to remember themselves of their tasks and focus.

**Debriefing phase:**
*Observations.* The students were surprised by how productive they were in such limited time. They expressed positive feedback about the strict time-frame of the working sessions which forced them to focus on the essential parts of their ideas. They suggested that the workshop should take place at the beginning of the project rather than half-way through it.
*Lessons learned 4.* Although the students experienced an unusual approach in comparison to their previous studies, they were fond of the workshop and engaged with the activities. They suggested.
*Suggestions 4.* Implementing the Design Thinking workshop not just as an intervention but rather as a regular activity at the beginning of the project.

## 4.2 AT THE END OF THE PROJECT.

In this section, we focus on the feedback obtained by the telecommunications team.

*Observation* The team had four members. Only two of them participated in the workshop. It became apparent that their teammates did not want to engage with the ideas that emerged during the workshop. They did not perceive the workshop as useful because it did not have an *applied* goal and it was *just* about creativity. However, after the customer expressed dissatisfaction with the team's technically-motivated ideas (e.g., a chatbot to improve customer support), they were motivated to work on the ideas which were the output of the workshop.
*Lessons learned 5.* Acceptance is critical to implement new ideas. The students need to understand and experience the ideation to implement the ideas.
*Suggestions 5.* All members of a team must be involved in the workshop so that each of them has the chance to contribute with her or his idea(s).

*Observation.* The team made some research afterwards and realized that most of the ideas generated during the workshop were already implemented by the customer.
*Lessons learned 6.* While the students were half way through the project, they were still not able to get an overview of the products/services of their customer. This can be due to improper research by the students or/and due to the big size of the company.

*Suggestions 6.* Involve the customer in the workshop and/or ask them for validation (e.g., *Test mode*) to get faster feedback and to avoid redundant ideas. Planning an "Ice-Breaker"—i.e., a game to prevent biased behaviour due to the different roles (e.g., customer, student, teacher).

*Observation.* According to the team, their final app idea did not originate from the workshop. However, the facilitator documented a quite similar idea during the first prototype phase.

*Lessons learned 7.* Students can have difficulty in accepting that their project idea originated from a context in which also student from other teams were involved. This indicates low acceptance towards shared ownerships of ideas.

*Suggestions 7.* In case of an *intervention* workshop, the facilitator should ask the students beforehand about their ideas and experiences to avoid redundancy. To that end, the facilitator can prepare an *ideation phase* for every team to increase the acceptance towards shared ownerships of ideas. Those side effects of projects-based learning (e.g., a team crisis) should be acknowledged to the students as learning opportunities which can be solved using Design Thinking.

## 5 CONCLUSION

In this paper, we reported our experience in running a workshop to engage teams of students involved in project-based learning using Design Thinking. In particular, we used the modes and mindsets proposed by the Stanford d.school to implement Design Thinking activities. After the workshop, the students got a human-centered perspective towards innovative solution generation. Moreover, we propose improvement for further iteration of the workshop based on our lessons learned.

## REFERENCES

[1] S. Kurbanoğlu and B. Akkoyunlu, "Information Literacy and Flipped Learning," in *In Pathways Into Information Literacy and Communities of Practice*, Chandos Publishing, 2017, pp. 53-84.

[2] K. Ivy, "A Flipped Classroom: Bridging Learning Opportunities Between the Classroom and the Real World," *E-Learn: World Conference on E-Learning in Corporate, Government, Healthcare, and Higher Education*, pp. 1787-1795, October 2015.

[3] A. Scheer and C. Noweski, "Transforming constructivist learning into action: Design thinking in education"," C.C.D.A.T. Education, 2012. [Online]. Available: ojs.lboro.ac.uk.

[4] I. Rauth, E. Köppen, B. Jobst and C. Meinel, "Design Thinking: An Educational Model towards Creative Confidence," in *DS 66-2: Proceeding of the 1st International Conference on Design Creativity*, 2010.

[5] T. Lindberg, C. Noweski and C. Meinel, "Evolving discourses on design thinking: how design cognition inspired meta-disciplinary creative collaboration," *Technoetic Arts: A Journal of Speculative Research*, no. 8(1).

[6] U. Johansson-Sköldberg, J. Woodilla and M. Cetinkaya, "Design Thinking: Past, Present and Possible Futures," *Creativity and Innovation Management*, vol. 22, no. 2, pp. 121-146, March 2013.

[7] C.L. Dym, A.M. Agogino, O. Eris, D.D.Frey and L.J.Leifer, "Engineering Design Thinking, Teaching and Learning," *Journal of Engineering Education*, vol. 94, no. 1, pp. 103-120, January 2005.

[8] Ju, W., Neeley Jr, W. L., & Leifer, L. J., "Design, design & design: an overview of Stanford's Center for Design Research," in *Conference on Human Factors in Computing Systems*, 2007.

[9] T. Seitz, Design Thinking und der neue Geist des Kapitalismus, Bielefeld: Transcript, 2017.

[10] L. Kimbell, "Rethinking Design Thinking: Part I," *Design and Culture*, vol. 3, pp. 285-306, 2011.

[11] d.school, "The Bootcamp Bootleg," [Online]. Available: https://dschool.stanford.edu/resources/the-bootcamp-bootleg. [Accessed 18 November 2017].

[12] d.school, "How to Kick Off a Crash Course," [Online]. Available: https://static1.squarespace.com/static/57c6b79629687fde090a0fdd/t/589931fdebbd1a3eaa99f495/1486434814313/d.mindsets_8.5-x-11_scissors.pdf. [Accessed 20 November 2017].

[13] H. Schuler and Y. Görlich, "Förderung durch organisationale Bedingungen," in *Kreativität*, Göttingen, Hogrefe Verlag GmbH & Co.KG, 2007, pp. 99-102.

[14] S. Gloger, "Arbeiten beim Kaffeetrinken," *managerSeminare*, vol. 75, pp. 50-56, 2004.